# Ultrafast, reconfigurable all-optical beam steering and spatial light modulation

Claudio U. Hail[1,2], Lior Michaeli[1,3], Harry A. Atwater[1]*

[1] Thomas J. Watson Laboratory of Applied Physics, California Institute of Technology, Pasadena, California 91125
[2] Department of Mechanical Engineering, University of California, Berkeley, California 94705
[3] School of Electrical and Computer Engineering, Faculty of Engineering, Tel-Aviv University, Tel-Aviv 6997801
* Correspondence and requests for materials should be addressed to H.A.A (email: haa@caltech.edu).

## Abstract

Achieving spatiotemporal control of light at subwavelength and subcycle scales is an important milestone in the development of new photonic materials and technologies. Ultrafast spatiotemporal light modulation currently relies on electronic interband and intraband transitions that yield pronounced refractive index changes but typically suffer from slow, picosecond response times due to carrier relaxation. Here we show that by leveraging resonant light-matter interactions in a high-quality factor metasurface it is possible to use the optical Kerr effect, a weaker, but instantaneous optoelectronic polarization effect, to achieve ultrafast, reconfigurable light modulation with unprecedented spatial and temporal control. By the subwavelength all-optical tuning of the refractive index of the dielectric metasurface unit cells, we experimentally demonstrate pulse-limited beam steering with a 74-fs response time at angles up to ±13° in the near-infrared. The steering originates from the Kerr effect with a background contribution arising from slower two-photon-excited free carrier absorption. Additionally, we observe spatial back-action, linear frequency conversion, and demonstrate arbitrary ultrafast spatial light modulation in two dimensions. Our findings open the possibility of realizing new ultrafast physics in metastructures with applications in signal processing, pulse shaping, and ultrafast imaging.

## Introduction

The ability to spatially manipulate light at wavelength and subwavelength scales with nanostructured materials has profoundly reshaped the way we think about optical materials and devices[1]. This precise spatial control over light lies at the center of photonic crystals, metamaterials, and metasurfaces that have enabled a host of advances in the past three decades, ranging from spontaneous emission control[2] to efficient flat lenses[3,4] and much more[5]. In an analogous manner, the control over light in *time*, at the timescale of a single optical cycle is expected to lead to numerous impactful developments such as optical time crystals[6–9], optical non-reciprocity[10,11], temporal antireflection coatings[12] or analog Hawking radiation[13]. So far, temporal control over light in nanoscale metastructures has been explored, among others, with mechanical[14–16], phase-change[17,18], liquid crystal[19,20], thermo-optic[21–23], or electro-optic methods[24–26], enabling key advancements such as GHz free-space light modulation[26] or dynamic varifocal lensing[24]. However, in most cases, changes in optical properties are quasistatic, i.e.

changes occur on a timescale much longer compared to the optical cycle. This is because in the optical range, the magnitude of refractive index changes trades off against the modulation speed, so that, with a few notable exceptions, ultrafast changes are typically vanishingly small. Only recently, by leveraging nanoengineered materials and exploring materials with enhanced nonlinearity, has it been possible to attain temporal modulation at optical frequencies to observe ultrafast time-varying material behavior[27–38].

Exploiting optical nonlinearities in semiconductor-based metasurfaces is a prominent method for the ultrafast modulation of optical material properties. Metasurfaces exhibiting high-quality factor (Q) optical resonances are particularly favorable due to the amplified electric field-induced changes in refractive index at resonance, and the high sensitivity of the optical resonance mode to these changes. Using semiconductor-based metasurfaces, it has been possible to realize ultrafast phenomena such as all-optical switching[39–41], photon acceleration[28], spectral pump self-action[31], frequency conversion[29,32], or incoherent emission steering[42]. More recently, spatiotemporal light control was also explored with a picosecond time response from the timescale of the photogenerated carriers[43–46]. Alternatively, the use of transparent conductive oxides (TCOs) is another promising route for attaining ultrafast light modulation due to their strong nonlinear response and the ability to operate at the epsilon-near-zero (ENZ) point[47,48], having enabled demonstrations of time refraction[30] and diffraction[36], and spatiotemporal light control[27,37,38]. However, despite much progress with semiconductor-based metasurfaces and TCOs, the spatiotemporal control over light in these materials is still limited with respect to ultrafast beam steering and spatial light modulation. New materials and structures are needed to further expand spatiotemporal control into the reconfigurable, subwavelength, and subcycle regime.

Here, we study the ultrafast, all-optical spatiotemporal wavefront shaping of a probe beam by leveraging the spatially structured optical pumping of a dielectric high-Q metasurface. We experimentally demonstrate reconfigurable, pulse-limited beam steering with a 74-fs response time at angles up to ±13°, by exploiting the all-optical tuning of subwavelength, higher-order Mie-resonant nanopillars with high quality factor. The beam steering and light modulation originate from the instantaneous optical Kerr effect, with a slower background contribution due to the free carrier generation via two-photon absorption. Additionally, we observe signatures of ultrafast spatiotemporal back-action and frequency conversion, and show ultrafast all-optical spatial light modulation.

**Results**

Figure 1a illustrates our scheme and metasurface platform for ultrafast all-optical spatial light modulation. We leverage the third-order nonlinearity of amorphous silicon (a-Si), giving rise to ultrafast all-optical changes in the refractive index of the material. In a-Si, these changes are primarily induced by the optical Kerr effect, a lossless and instantaneous response arising from the electronic polarization of the material, and the free carrier effect, which is lossy and shows a finite rise time and picosecond decay time due to carrier relaxation[49]. These effects are captured by the intensity and time-dependent refractive index $n(t,I) = n_0 + n_2 I + \Delta n_{FC}(I,t)$, where $I$ represents the illuminated light intensity, $n_2$ the nonlinear refractive index due to the Kerr effect, and $\Delta n_{FC}(I,t)$ accounts for the dynamic, complex refractive index modification from photoexcited carriers. A main challenge is that these refractive index changes are typically very small, for example $n_2$ = 0.0002 cm$^2$/GW[50], and for peak pump intensities of 10 GW/cm$^2$ the free carrier contribution can reach Re[$\Delta n_{FC}$] ~ - 0.02 and Im[$\Delta n_{FC}$] ~ - 0.01[39]. By structuring the a-Si thin film into a high-Q optical metasurface, we leverage these small refractive index changes to induce large, ultrafast, all-optical changes in the optical response of the material. The metasurface is composed of a sub-diffractive array of a-Si nanopillars of length *L,* height *H* and periodicity *P*, on a transparent glass substrate. These geometrical dimensions are designed to induce higher-order Mie-modes that are known to generate a local high-Q response at the design wavelength in the near-infrared spectral range[51]. This type of metasurface shows strong field enhancement, low angular dispersion, and allows for wavefront shaping[51], enhanced third-harmonic generation[52] or thermo-optic beam steering[23]. For ultrafast light modulation, we adopt a degenerate pump-probe configuration, where the transmission of a normally-incident weak (probe) pulse is modulated by a strong (pump) pulse at the same wavelength. Due to the optical Kerr effect and the free carrier generation, the a-Si experiences ultrafast spatiotemporal pump-induced changes in the refractive index, *n(x,y,z,t)*, enabling the ultrafast modulation of the probe pulse in space and time.

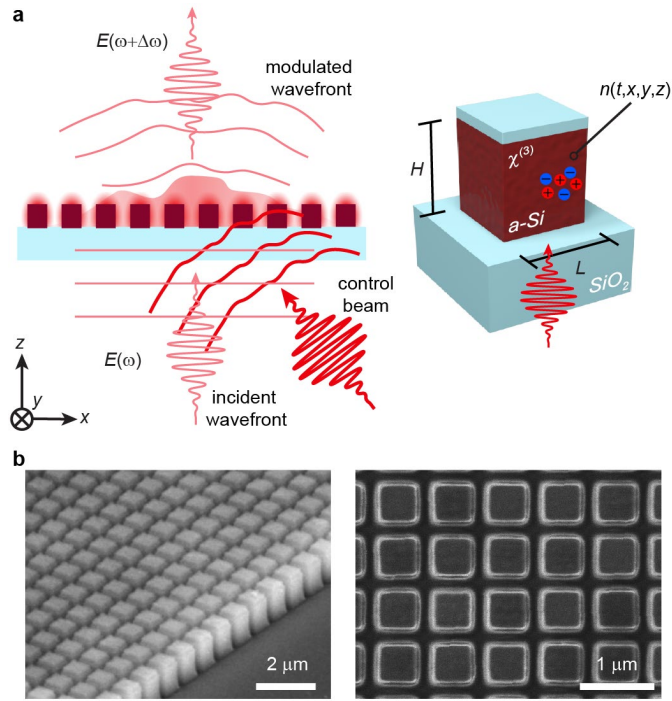

**Figure 1. | High-Q metasurfaces for ultrafast all-optical wavefront manipulation. a,** Schematic of the metasurface where the wavefront of a weak beam (probe) is modulated in transmission by a spatially varying control beam (pump) of the same wavelength, and a schematic of the corresponding metasurface unit cell composed of amorphous silicon square pillars with length *L* and height *H*. **b,** Scanning electron micrographs of the metasurface from a tilted view and a top view.

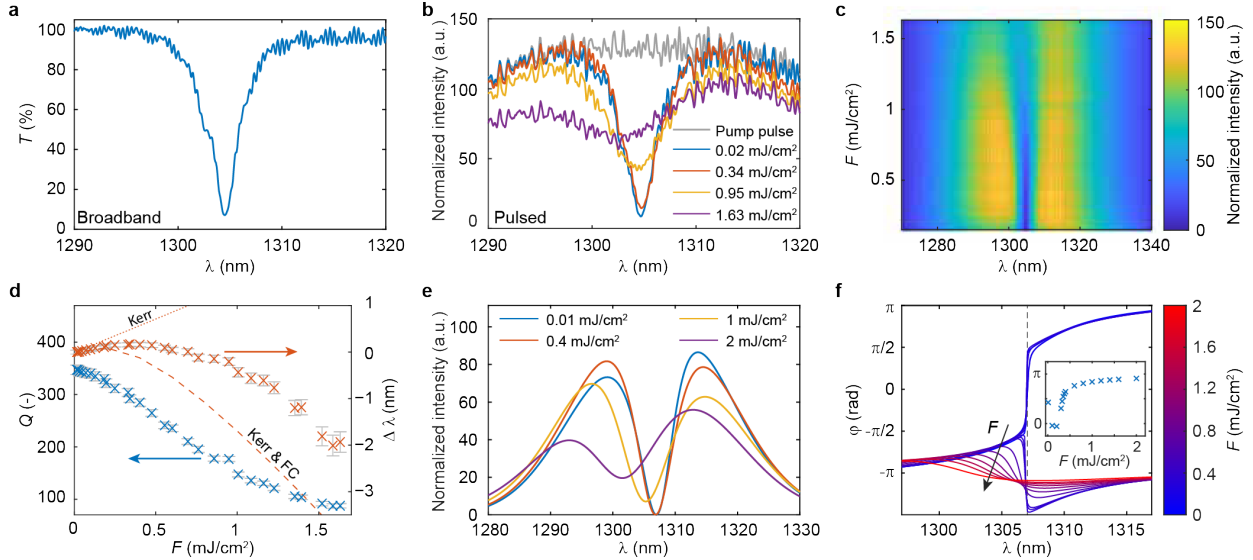

**Figure 2. | All-optical self-modulation on a high-Q metasurface. a**, Experimentally measured transmission spectrum of the metasurface with $P$ = 736 nm, $H$ = 695 nm, $L$ = 604 nm under low-power broadband illumination with a supercontinuum source. **b**, and **c**, Experimentally measured fluence-dependent normalized intensity spectrum of the light transmitted through the metasurface in (**a**) under resonant illumination with a fs-pulsed laser (100 fs, 1 kHz) with varying pump fluence. The spectra are normalized by the indicated incident energy density. **d**, Experimentally measured quality factors (blue) and change in resonance wavelength (orange) with varying pump fluence (crosses) extracted from measurements in (**c**). Simulated change in resonance wavelength considering the optical Kerr and free-carrier effects (dashed), and only the optical Kerr effect (dotted). The error bars represent the 95% confidence interval from the Fano fit to the spectrum. **e**, Simulated normalized intensity spectrum of the light transmitted through the metasurface in (**a**) under resonant illumination with a 100-fs laser pulse with varying pump fluence. **f**, Simulated change in spectral phase of the transmitted light with varying pump fluence. The inset shows phase values at 1307.3 nm with the values wrapped between [-π,π].

To study ultrafast light modulation, we fabricated the metasurfaces and first characterized their linear and nonlinear optical response under uniform illumination. The surfaces were fabricated in a cleanroom environment with a single-step electron beam lithography and dry etching process on plasma-deposited thin films of a-Si and silicon dioxide (see Methods). Figure 1b illustrates scanning electron microscopy images of a metasurface for ultrafast spatial light modulation. For the linear optical characterization, we illuminated the surface at normal incidence with light from a low power supercontinuum source and recorded the spectral distribution of the transmitted light on a grating spectrometer. Figure 2a shows the measured transmission spectrum of the metasurface exhibiting a narrow transmission dip at the resonance wavelength $\lambda_0$ = 1304.5 nm with $Q$ = 370, in good agreement with previous work[51]. This results in a resonator lifetime of the light in the nanopillar of $\tau_M = 2Q/\omega_0$ = 513 fs. For the nonlinear optical characterization, we resonantly illuminated the surface at normal incidence with laser pulses from a Ti:Sapphire amplifier (pulse length $\tau_p$ = 100 fs, repetition rate 10 kHz) that were wavelength-converted by an optical parametric amplifier, and recorded the spectrum of the transmitted pulses. To avoid interpulse effects, a chopper reduced the laser pulse repetition rate to 1 kHz. Figures 2b and 2c illustrate the measured power-dependent transmitted spectral intensity of the metasurface, normalized by the respective pump fluence. Both the measured resonance wavelength and quality factor are power dependent, as illustrated with the extracted values in Fig. 2d. For $F$ <

0.05 mJ/cm$^2$, $\lambda_0$ and $Q$ are nearly identical to the linear response. With increasing fluence, we observe first a spectral redshift of up to $\Delta\lambda$ = 0.17 nm, and then a blueshift with strong broadening of the resonance up to $\Delta\lambda$ = -2 nm and $Q$ = 87 for $F$ = 1.59 mJ/cm$^2$. Measurements of different metasurfaces with varying $L$ and $\lambda_0$ show a similar behavior (see Supplementary Figs. 1 and 2). The observed spectral redshift indicates an increase in the refractive index consistent with the instantaneous Kerr effect (thermo-optic effects can be neglected at this repetition rate), while the blueshift and spectral broadening point to a decrease in refractive index associated with the Drude response of the photoexcited free carriers. To provide further insight into these nonlinear properties of our metasurface, we performed time-domain simulations of Maxwell's equations using a finite element multiphysics solver. The model incorporates a spatially and temporally varying refractive index that includes contributions from the optical Kerr effect, two-photon absorption, the free-carrier population dynamics via a rate equation, Drude dispersion, and the optical response of the free carriers (see Methods and Supplementary Note 1). Figure 2e illustrates the simulated normalized transmitted spectral light intensity of the metasurface for varying pump fluence, showing a qualitatively similar response as the corresponding measurement in Fig. 2b. The simulated shift in resonance wavelength is presented in Fig. 2d, comparing the response of the combined free-carrier and Kerr effect with the Kerr effect alone. These results show that the redshift at low fluence originates from the optical Kerr effect, and the blueshift and broadening at larger fluence from the free carrier generation. Finally, we extract the power-dependent phase of the transmitted light from the simulations (Fig. 2f). With increasing power, a phase shift of approximately 180° is obtained on resonance (see inset), primarily limited by increased absorption from the free carrier response. This suggests that a degenerate pump beam can modulate the amplitude and phase of the transmitted wavefront of an incident probe beam.

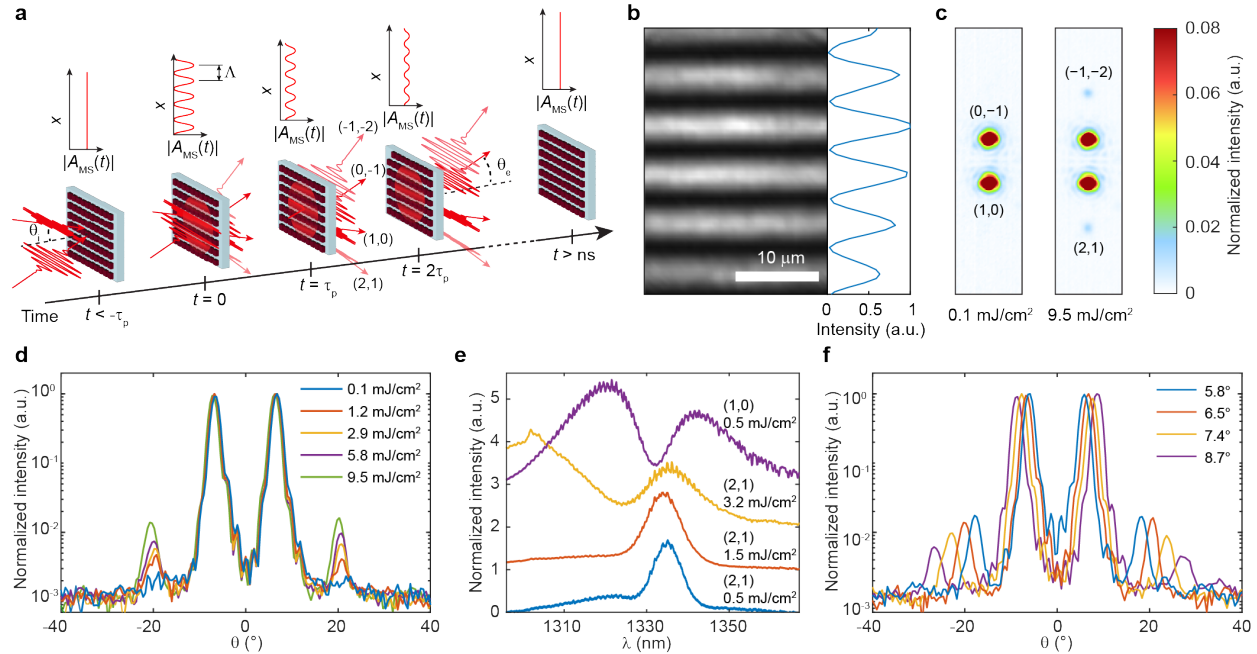

**Figure 3. | Spatiotemporal self-diffraction on a transient optical grating. a**, Schematic of the metasurface illuminated with two pump beams at angles ± $\theta_i$, forming a fluence-dependent transient optical grating with a spatiotemporally varying transmission amplitude $A_{MS}$. This results in the diffraction of the pump on the grating that was formed by the pump itself to diffraction orders (-1,-2), (0,-1), (1,0), and (2,1).

**b**, Experimentally measured electric field intensity at the metasurface plane for illumination with $\theta_i = \pm 6.5°$. The line plot represents a cross section through the center. The pump illumination is on resonance at $\lambda = 1332$ nm with a 100-fs pulse width and 1 kHz repetition rate. **c**, Experimentally measured images of the Fourier plane of the light transmitted through the metasurface for illumination with $\theta_i = \pm 6.5°$ and $F = 0.1$ mJ/cm$^2$ and $F = 9.5$ mJ/cm$^2$. **d**, Cross-sectional line cuts of experimentally measured images of the Fourier plane of the transmitted light for illumination with $\theta_i = \pm 6.5°$ and varying pump fluence. **e**, Experimentally measured spectrum of normalized intensity of the light transmitted to the (2,1) diffraction order for varying pump fluence and to the (1,0) diffraction order for reference. Spectra are shifted vertically by an increment of 1 for better visibility. **f**, Cross-sectional line cuts of experimentally measured images of the Fourier plane of the transmitted light for illumination with varying incidence angles $\theta_i$ and a pump fluence of approximately $F = 9$ mJ/cm$^2$.

As a first characterization of the spatiotemporal light modulation, we illuminate the metasurface with a transient optical grating and analyze the angular distribution of the transmitted light. As illustrated in Fig. 3a, two pump beams simultaneously illuminate the metasurface at angles $\pm\theta_i$ and interfere to form an optical grating with a period of $\Lambda = \lambda/(2\sin\theta_i)$ on the metasurface. This induces a spatiotemporally varying refractive index that breaks the uniformity of the metasurface. In the experiment, the surface is resonantly illuminated with $\theta_i = \pm 6.5°$ and the transmitted pump beams are collected with an objective lens and imaged onto an InGaAs camera. Here, a metasurface with $L = 619$ nm and $\lambda_0 = 1332$ nm is measured. Figure 3b shows a measured image of the optical grating that is formed at the metasurface plane with $\Lambda = 5.88$ μm. Figure 3c displays the measured Fourier plane of the transmitted pump at $F = 0.1$ mJ/cm$^2$ and at $F = 9.5$ mJ/cm$^2$. To label the various diffracted transmitted beams, we use the notation $(m_+, m_-)$, where $m_+$ and $m_-$ denote the diffraction order relative to the beam incident at $+\theta_i$ and $-\theta_i$, respectively. Due to the symmetric illumination, the diffraction orders are degenerate with $m_+ = m_- + 1$. At low fluence, the metasurface exhibits a linear response with negligible refractive index modulation, resulting in the specular transmission of the incident pulses to (0,-1) and (1,0) as they impinge on an effectively uniform material. Notably, with increasing fluence, additional, higher diffraction orders (-1,-2) and (2,1) appear in the Fourier plane at angles $\theta_e = \pm 20°$. A cross-section across the Fourier plane reveals that the light intensity deflected into these higher diffraction orders monotonously increases with pump fluence (see Fig. 3d). This shows that the pump experiences spatial self-modulation as it diffracts on itself due to the all-optical spatiotemporal modulation of the refractive index of the metasurface. As the preceding part of the pump pulse induces the modulation, the succeeding part deflects on the transient grating. As a reference, the same measurement on a thin film of the identical material stack shows no diffraction orders (-1,-2) and (2,1) even at $F = 9.5$ mJ/cm$^2$ (see Supplementary Fig. S3). This shows that the high-Q metasurface is instrumental in achieving ultrafast spatiotemporal light modulation. Figure 3e illustrates the spectral distribution of the light diffracted to the specular and higher diffraction orders at varying pump fluences. The specular transmitted beam (0,1) shows a transmission dip due to the high-Q resonance of the metasurface. Up to a fluence of 1.5 mJ/cm$^2$, the deflected beam (1,2) is mainly diffracted near the resonance wavelength of the metasurface. For $F = 3.2$ mJ/cm$^2$, the light is partly diffracted on- and off-resonance with a considerable part being blue shifted. This suggests that at low pump fluence the light deflection mainly occurs through phase modulation of the high-Q resonance, while at increased fluence a significant portion of the incident light is deflected off-resonance due to amplitude modulation from increased losses of free carrier effects. By varying the incident angle $\theta_i$, and thus $\Lambda$, light can be deflected to varying angles. This is shown in the measured cross-sections through the Fourier plane in Fig. 3f (see also Supplementary Figs. 4-6).

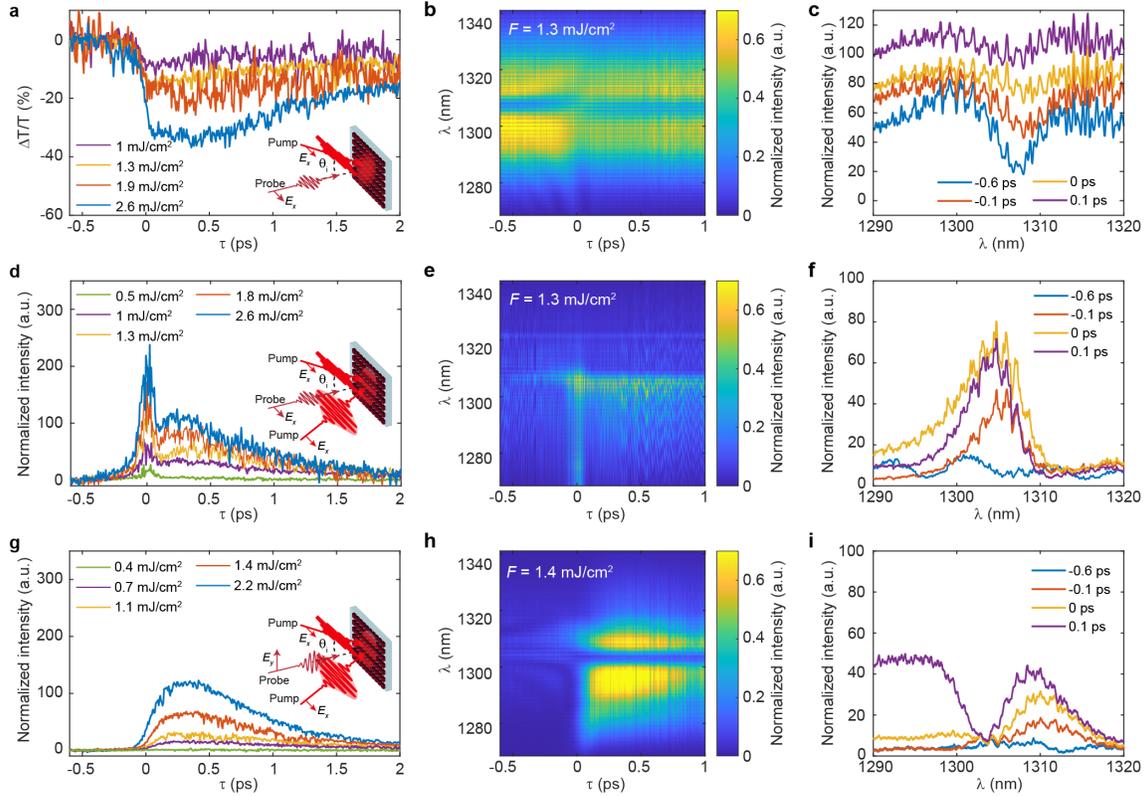

**Figure 4. | Time-dependent analysis of a metasurface for ultrafast all-optical beam deflection. a,** Experimentally measured differential transmission of the resonant probe beam on the metasurface with varying pump-probe delay times, for a uniform pump illumination at $\theta_i = \pm 6.4°$ and varying pump fluence. Pump and probe beams are co-polarized. **b**, **c**, Spectral intensity of transmitted probe beam for the measurement in (**a**) at $F = 1.3$ mJ/cm$^2$ and varying pump-probe delay times. The spectra in (**c**) are shifted with increasing $\tau$ by 20% for better visibility. **d,** Experimentally measured normalized intensity of the probe beam transmitted to the (2,1) diffraction order with varying pump-probe delay times, for grating illumination at $\theta_i = \pm 6.4°$ and varying pump fluence. **e**, **f**, Spectral intensity of the light transmitted to the (2,1) diffraction order the measurement in (**d**) for $F = 1.3$ mJ/cm$^2$ and varying pump-probe delay times. Pump and probe beams are co-polarized. **g,** Experimentally measured normalized intensity of the probe beam transmitted to the (2,1) diffraction order with varying pump-probe delay times, for a grating illumination at $\theta_i = \pm 6.4°$ and varying pump fluence. Pump and probe beams are cross-polarized. **h**, **i**, Spectral intensity of the light transmitted to the (2,1) diffraction order the measurement in (**g**) for $F = 1.4$ mJ/cm$^2$ and varying pump-probe delay times. The insets in (a), (d), and (g) illustrate the illumination conditions and polarization of pump and probe beams for each row of measurements.

To provide insight into the temporal aspect of the ultrafast beam modulation, we performed time-resolved transmission spectroscopy in a degenerate pump probe configuration on the metasurface. For this measurement, a metasurface with $L = 604$ nm and $\lambda_0 = 1304.5$ nm was illuminated with a single pump beam at an angle of $\theta_i = 6.4°$ and a normally-incident, co-polarized probe beam. Figure 4a illustrates the measured transient pump power-dependent differential transmission at different time delays $\tau$ between the arrival of the pump and probe, where a negative time delay indicates that the probe arrives before the pump beam. At $\tau > 0$, a strong decrease in transmission is observed with a maximum of -36% for the maximum fluence. From an exponential fit to the transmission, relaxation times ranging between 2.3 – 3.4 ps are determined, which are typical for free carrier recombination in amorphous silicon[53]. The corresponding measured spectral normalized transmitted light intensity for varying pump-probe delays $\tau$ at a pump fluence of $F$ = 1.3 mJ/cm$^2$ is shown in Fig. 4b and 4c. At negative delays longer than the residence time of the photons, i.e. $\tau < -\tau_M$, there is no effect of the pump on the probe, and the unperturbed transmission spectrum is observed. At $\tau_M < \tau < 0$, a clear redshift of the metasurface resonance is observed, and then at $\tau > 0$ the resonance is blue shifted. This is a manifestation of the Kerr effect and the free carrier effect acting at $\tau_M < \tau < 0$ and $\tau > 0$ timescales, respectively. Importantly, the interference fringes at $\tau < 0$ and $\lambda < 1300$ nm, show that light is undergoing linear frequency conversion due to the ultrafast change in refractive index, as previously observed in high-Q silicon-based metasurfaces[29]. Similar observations are made at other fluence values (see Supplementary Fig. 7). This interference pattern results from interference of probe light which was coupled to the resonator and frequency-shifted by the pump beam, with residual probe light that did not interact with the resonator[29].

To study the ultrafast beam deflection, a second pump beam is added to illuminate the surface at $\theta_i = \pm 6.4°$ with a transient optical grating. Figure 4d shows the light intensity deflected to the transmitted +1 diffraction order for a normally-incident, co-polarized probe beam at different pump powers. An ultrafast peak is observed at $\tau = 0$ with a sharp rise of the diffracted light intensity at $\tau < 0$ and a steep drop at $\tau > 0$. The duration of the peak ranges between 74–90 fs for varying fluence, which is of similar length as the pulse length of the pump determined by linear autocorrelation (see Supplementary Figs. 8 and 9). A second broader peak at $\tau \approx 0.3$ ps is observed with a decay time of 0.7-1.2 ps, which we attribute to the response of the photoexcited carriers. These measurements reveal two distinct contributions to the beam deflection: an ultrafast, instantaneous component driven by the Kerr effect, and a weaker and slower background arising from free carrier dynamics. This interpretation is supported by simulations of the time-dependent refractive index evolution (see Supplementary Fig. 15), which reproduce both timescales observed experimentally. Notably, the simulations predict all-optical, pulse-limited changes in refractive index of up to $\Delta n = 0.018$ (corresponding to $n_2$ = 0.0018 cm$^2$/GW) at the center of the nanopillar, which is on par with previously reported refractive index changes due to photocarrier generation[39]. The corresponding normalized measured spectral light intensity of the transmitted +1 diffraction order of the probe beam for $F$ = 1.3 mJ/cm$^2$ is shown in Figs. 4e and 4f (see Supplementary Fig. 10 for spectra at other fluence). At $\tau < -\tau_M$, there is no light deflection as probe light is not impacted by the structured pump beam. At $-\tau_M < \tau < 0$, light is deflected at the resonance wavelength, and at $0 < \tau$ the deflected light blueshifts. An identical reference measurement with a co-polarized probe on a thin film shows no beam deflection at any time delays (see Supplementary Fig. 11). This further underscores the key role of the high-Q metasurface for the ultrafast light modulation. As an additional control measurement, we performed the identical experiment with a cross-polarized probe at normal incidence (Figs. 4g, 4h, 4i and Supplementary Fig. 12). In this case, only light deflection from free carriers is observed without the Kerr component, since the Kerr effect only affects the refractive index along the direction of the electric field. These results show that our high-Q metasurface enables ultrafast all-optical beam deflection with pulse-limited performance driven by the optical Kerr effect accompanied by a slower background free carrier contribution.

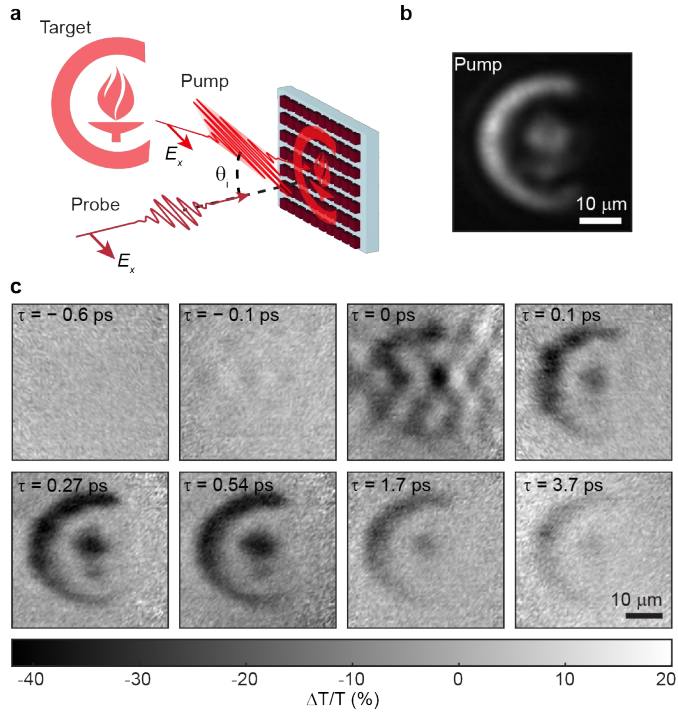

**Figure 5. | Ultrafast all-optical spatial light modulation. a,** Schematic of the measurement configuration where the pump beam projects an arbitrary image, here the Caltech logo as an example, onto the metasurface. **b**, Experimentally measured intensity of the image projection of the transmitted resonant pump beam at the metasurface plane. The pump illumination is at an angle of θ = 9.1° on resonance at λ = 1307 nm with a 100-fs pulse width and 1 kHz repetition rate. **c**) Experimentally measured differential transmission images of the resonant probe beam at varying pump-probe time delays. Interference effects at τ = -0.1, 0 and 0.1 ps are due to interference of non-resonant probe light with frequency-shifted probe light.

      To illustrate the full potential of our modulation mechanism, we spatiotemporally modulate the probe beam with an arbitrary two-dimensional pattern. As illustrated in Fig. 5a, a pump beam with a target pattern is projected onto the metasurface at $\theta_i$ = 9.1°. The probe beam is normally incident and co-polarized to the pump beam. Figure 5b illustrates a measured image of the pump beam projected onto the metasurface. Acquired differential transmission images $\Delta T(\tau, x, y)/T(x, y)$ of the probe beam at varying pump-probe delay times τ are shown in Fig. 5c. For $\tau < -\tau_M$, the transmission of the probe beam is unperturbed. For τ ≥ 0.27 ps, the pump pattern is clearly visible in the differential transmission of the probe. The observed ultrafast imaging shows a spatial resolution of 3 μm, corresponding to the smallest gap on the bottom between the letter "C" and the torch above. Here, the resolution is limited by the projection optics (numerical aperture 0.25) that fail to resolve the smaller features of the torch. With longer time delays, the pattern fades in intensity, as expected from the free carrier relaxation. Notably, at time delays around τ = 0 ps, a pronounced interference pattern is observed in the differential transmission of the probe. We hypothesize that this is interference of probe light that was frequency-converted with remaining probe light that did not interact with the metasurface, as a spatial manifestation of the interference observed in the spectra in Fig. 4b.

The modulation scheme presented here is reconfigurable, and its optical function (beam steering, spatial modulation, lensing, etc.) entirely depends on the pump intensity pattern. To the best of our knowledge, this is the first demonstration of fully reconfigurable pulse-limited beam steering and spatial light modulation of specular light at the nanoscale. With an effectively instant response and high spatial resolution of modulation, our results significantly expand the range of both spatial and temporal modulation compared to the currently available methods for ultrafast nanoscale light manipulation. Other similar demonstrations have been limited to binary switching at a slower timescale[43], limited reconfigurability[27], beam switching of upconverted light[54,55], or show a small angular range of less than 1.5°[34,38,56]. Notably, our realization of spatial light modulation may also serve as an ultrafast display or camera for imaging. With better projection optics, the spatial modulation resolution may approach the metasurface unit cell size, allowing deflection of light to steep angles. While the high Q factors currently limit the operation bandwidth, structures with Q boosting[57,58] could offer more broadband light coupling. Additionally, the effects of two-photon absorption may be reduced at different operating wavelengths or by employing spectral mode engineering[39].

**Conclusion**

In conclusion, we demonstrated the ultrafast, reconfigurable spatiotemporal steering and structuring of light at the nanoscale by spatially selective optical pumping of a dielectric higher-order Mie-resonant metasurface. We leverage the optical Kerr effect and the free carrier effect to induce beam steering with pulse-limited time response at angles of up to ±13°. Additionally, by illuminating the metasurface with a transient optical grating we observe ultrafast spatial back-action of the pump beam. Notably, we show that our method can be expanded to ultrafast two-dimensional spatial modulation of light with near-diffraction-limited resolution. Our results suggest that high-Q semiconductor metasurfaces may prove to be a viable platform for ultrafast signal processing and computing, and the realization of optical Floquet metamaterials[59] and time crystals at optical frequencies[9].

**Methods**

**Experiment**

Experiments were performed on a custom-built optical microscope illustrated in Supplementary Fig. 13. Light from a Ti:Sapphire Amplifier (Coherent, Libra, 10 kHz, 100 fs) is frequency-converted to the NIR using an optical parametric amplifier (Coherent OPerA Solo), split into a pump and probe beam and then impinged on the metasurface sample. The probe beam is routed via a delay stage to control the time delay between pump and probe beams. An exchangeable mask in the excitation path of the pump beam allows for switching between different configurations of pump illumination: normal-incident illumination, oblique illumination, symmetric oblique illumination, or the projection of an arbitrary 2D pattern. Light transmitted through the metasurface is collected by an objective lens (0.95 NA) and subsequently sent to an InGaAs camera, a grating spectrometer, or a photodiode, depending on the measurement. For imaging and filtering, a 4-f configuration is employed in the detection path. For pump-probe measurements the probe fluence was set to 0.2 mJ/cm$^2$ and the pump beam was separated by spatial filtering in a Fourier plane or with a polarizer for cross-polarized pump-probe measurements. The quality factor of the metasurface was determined by fitting a Fano line shape to the transmission. The high frequency oscillations in the transmission signal are due to interference effects in the metasurface substrate and residual reflections of optical components in the experimental apparatus.

**Fabrication**

The metasurfaces were fabricated on borosilicate glass cover slips ($n$ = 1.503, thickness 220 μm). Thin films of amorphous silicon and silicon oxide were deposited using plasma-enhanced vapor deposition. Prior to the deposition, substrates were solvent and oxygen plasma cleaned. The metasurface structure was then written into a spin-coated resist layer (MaN-2403) using standard electron beam lithography (100 kV) with subsequent development in MF-319. A thin layer of evaporated chrome was used as a charge dissipation layer. The resist pattern was then transferred to the thin film using a combination of reactive ion etching and chlorine-based inductively coupled reactive ion etching. The remaining resist layer was then removed by oxygen plasma cleaning.

**Simulations**

The nonlinear optical response of the metasurfaces was simulated using a time-domain finite element method with a commercially available solver (COMSOL Multiphysics). To this end, Maxwell's equations, the charge carrier rate equation, and the carrier drift-diffusion equation were numerically solved, accounting for two-photon absorption, the optical Kerr effect and a Drude correction to the refractive index (see Supplementary Note 1 for more details). Perfect magnetic and electric conductor boundary conditions, coupled with periodic boundary conditions, were used along the *x* and *y* directions, respectively. Along the *z*-axis, perfectly matched layer boundary conditions were used. For illumination, a Gaussian pulse with a full-width half-maximum pulse length of 100 fs was injected into the simulation domain. For the glass, a constant refractive index of $n$ = 1.503 was adopted and $n$ = 3.45 for amorphous silicon.

10. Sounas, D. L. & Alù, A. Non-reciprocal photonics based on time modulation. *Nat. Photonics* **11**, 774–783 (2017).
11. Shaltout, A., Kildishev, A. & Shalaev, V. Time-varying metasurfaces and Lorentz non-reciprocity. **5**, 225–230 (2015).
12. Pacheco-Peña, V. & Engheta, N. Antireflection temporal coatings. *Optica* **7**, 323 (2020).
13. Rubino, E. *et al.* Experimental evidence of analogue Hawking radiation from ultrashort laser pulse filaments. *New J. Phys.* **13**, 085005 (2011).
14. Arbabi, E. *et al.* MEMS-tunable dielectric metasurface lens. *Nat. Commun.* **9**, (2018).
15. Meng, C. *et al.* Dynamic piezoelectric MEMS-based optical metasurfaces. *Sci. Adv.* **7**, (2021).
16. Afridi, A., Gieseler, J., Meyer, N. & Quidant, R. Ultrathin Tunable Optomechanical Metalens. *Nano Lett.* **23**, 2496–2501 (2023).
17. Wang, Y. *et al.* Electrical tuning of phase-change antennas and metasurfaces. *Nat. Nanotechnol.* **16**, 667–672 (2021).
18. Zhang, Y. *et al.* Electrically reconfigurable non-volatile metasurface using low-loss optical phase-change material. *Nat. Nanotechnol.* **16**, 661–666 (2021).
19. Sautter, J. *et al.* Active tuning of all-dielectric metasurfaces. *ACS Nano* **9**, 4308–4315 (2015).
20. Li, S. Q. *et al.* Phase-only transmissive spatial light modulator based on tunable dielectric metasurface. *Science* **364**, 1087–1090 (2019).
21. Malek, S. C., Overvig, A. C., Shrestha, S. & Yu, N. Active nonlocal metasurfaces. *Nanophotonics* **10**, 655–665 (2021).
22. Archetti, A. *et al.* Thermally reconfigurable metalens. *Nanophotonics* **11**, 3969–3980 (2022).
23. Sokhoyan, R., Hail, C. U., Foley, M., Grajower, M. & Atwater, H. A. All-dielectric high-Q dynamically tunable transmissive metasurfaces. *arXiv* (2023) doi:10.48550/arXiv.2309.08031.
24. Shirmanesh, G. K., Sokhoyan, R., Wu, P. C. & Atwater, H. A. Electro-optically Tunable Multifunctional Metasurfaces. *ACS Nano* **14**, 6912–6920 (2020).
25. Weiss, A. *et al.* Tunable Metasurface Using Thin-Film Lithium Niobate in the Telecom Regime. *ACS Photonics* **9**, 605–612 (2022).
26. Benea-Chelmus, I. C. *et al.* Gigahertz free-space electro-optic modulators based on Mie resonances. *Nat. Commun.* **13**, 1–9 (2022).
27. Shaltout, A. M. *et al.* Spatiotemporal light control with frequency-gradient metasurfaces. *Science* **365**, 374–377 (2019).
28. Shcherbakov, M. R. *et al.* Photon acceleration and tunable broadband harmonics generation in nonlinear time-dependent metasurfaces. *Nat. Commun.* **10**, 1–9 (2019).
29. Karl, N. *et al.* Frequency Conversion in a Time-Variant Dielectric Metasurface. *Nano Lett.* **20**, 7052–7058 (2020).
30. Zhou, Y. *et al.* Broadband frequency translation through time refraction in an epsilon-near-zero material. *Nat. Commun.* doi:10.1038/s41467-020-15682-2.
31. Sinev, I. S. *et al.* Observation of Ultrafast Self-Action Effects in Quasi-BIC Resonant Metasurfaces. *Nano Lett.* **21**, 8848–8855 (2021).
32. Zubyuk, V. V., Shafirin, P. A., Shcherbakov, M. R., Shvets, G. & Fedyanin, A. A. Externally Driven Nonlinear Time-Variant Metasurfaces. *ACS Photonics* **9**, 493–502 (2022).
Manuscript: Ultrafast, reconfigurable all-optical beam steering and spatial light modulation   Page **12** of **14**

**Acknowledgements**

This work was supported by the Air Force Office of Scientific Research under grant FA9550-18-1-0354 and the Meta-Imaging MURI grant #FA9550-21-1-0312. C.U.H. also acknowledges support from the Swiss National Science Foundation through the Early Postdoc Mobility Fellowship grant #P2EZP2_191880 and the Postdoc Mobility Grant #P500PT_214452. L.M. acknowledges support from the Fulbright Fellowship program and the Breakthrough Foundation. We gratefully acknowledge the critical support and infrastructure provided for this work by The Kavli Nanoscience Institute at Caltech.

**Author Contributions**

C.U.H., L.M. and H.A.A conceived the project. C.U.H. performed the simulations, fabricated the devices, built the experiment, performed the measurements, and analyzed the results. L.M. assisted in experimental design, simulations, data analysis and interpretation. C.U.H. wrote the manuscript with input from all other authors. H.A.A supervised all aspects of the project.

**Competing Interests**

The authors declare no competing financial interest.

**Correspondence and requests for materials** should be addressed to H.A.A.

# Supplementary Information:

# Ultrafast, reconfigurable all-optical beam steering and spatial light modulation

Claudio U. Hail[1,2], Lior Michaeli[1,3], Harry A. Atwater[1]*

[1] Thomas J. Watson Laboratory of Applied Physics, California Institute of Technology, Pasadena, California 91125
[2] Department of Mechanical Engineering, University of California, Berkeley, California, 94705
[3] School of Electrical and Computer Engineering, Faculty of Engineering, Tel-Aviv University, Tel-Aviv 6997801

## Table of Contents

## Supplementary Figures

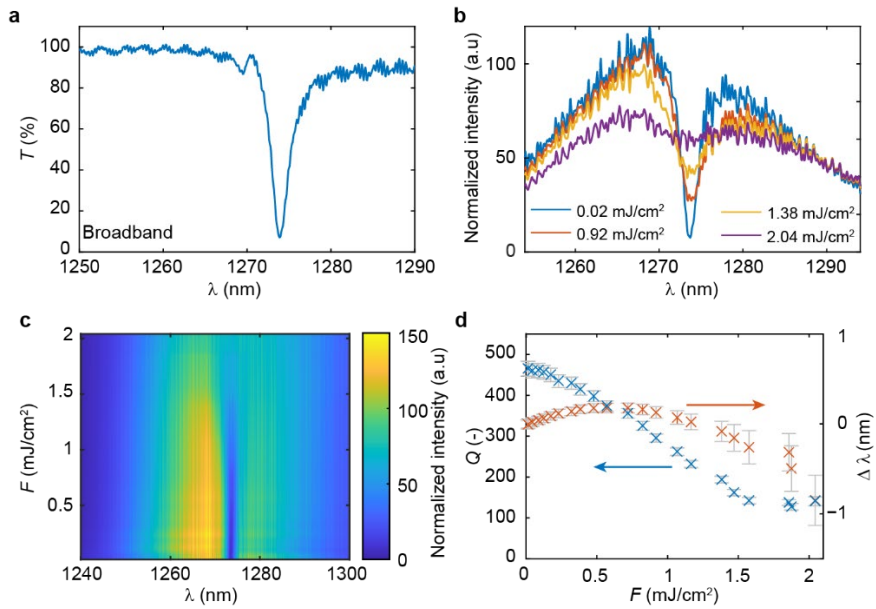

**Supplementary Figure 1 | All-optical self-modulation on a high-Q metasurface with *L* = 572 nm. a**, Experimentally measured transmission of a metasurface with *L* = 572 nm, *H* = 695 nm, and *P* = 736 nm under broadband illumination with a weak supercontinuum source. **b**, **c,** Experimentally measured normalized transmitted light intensity of the metasurface in (**a**) under pulsed illumination (100 fs, 1 kHz) on resonance for varying pump fluence. **d,** Experimentally measured resonance wavelength shifts and quality factors for varying pump fluence.

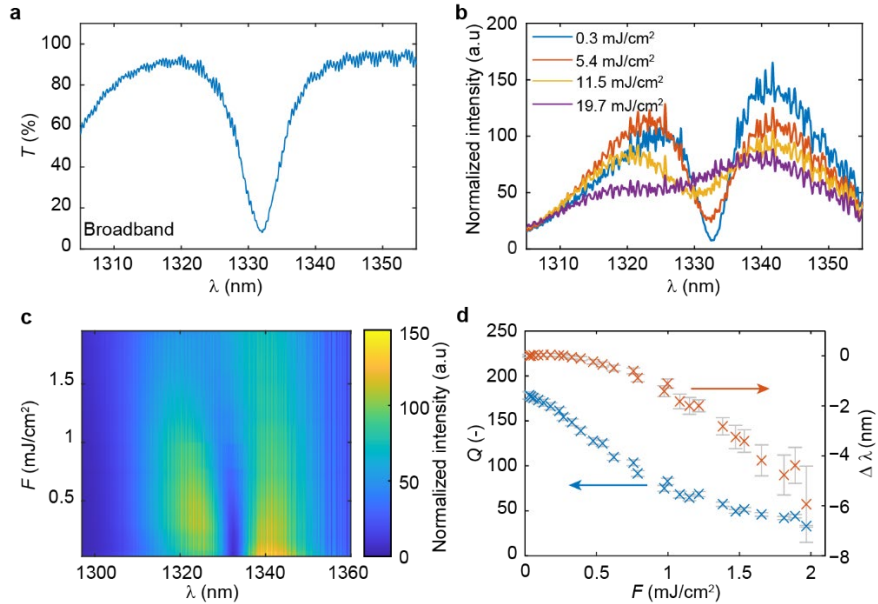

**Supplementary Figure 2 | All-optical self-modulation on a high-Q metasurface with *L* = 619 nm. a**, Experimentally measured transmission of a metasurface with $L$ = 619 nm, $H$ = 695 nm, and $P$ = 736 nm under broadband illumination with a weak supercontinuum source. **b**, **c,** Experimentally measured normalized transmitted light intensity of the metasurface in (**a**) under pulsed illumination (100 fs, 1 kHz) on resonance for varying pump fluence. **d,** Experimentally measured resonance wavelength shifts and quality factors for varying pump fluence.

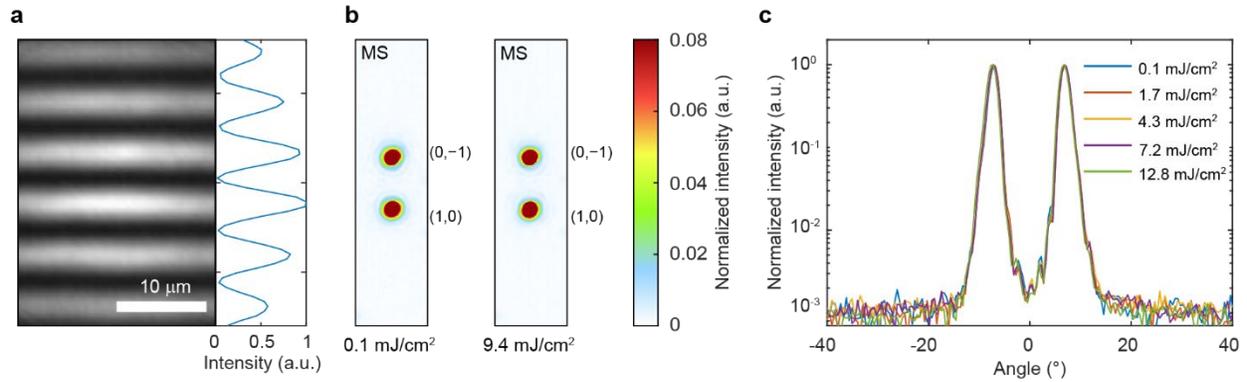

**Supplementary Figure 3 | Spatiotemporal self-diffraction reference measurement. a**, Experimentally measured light intensity at a reference thin film stack of a-Si and SiO$_2$ with an optical grating period of $\Lambda = 5.89$ μm. **b**, Experimentally measured Fourier plane images of the light transmitted through the thin film stack at pump fluence of $F = 0.1$ mJ/cm$^2$ and of $F = 9.4$ mJ/cm$^2$ including the labeled grating orders. **c**, Line cross-section of the experimentally measured Fourier plane images of the light transmitted through the thin film stack at varying pump fluence.

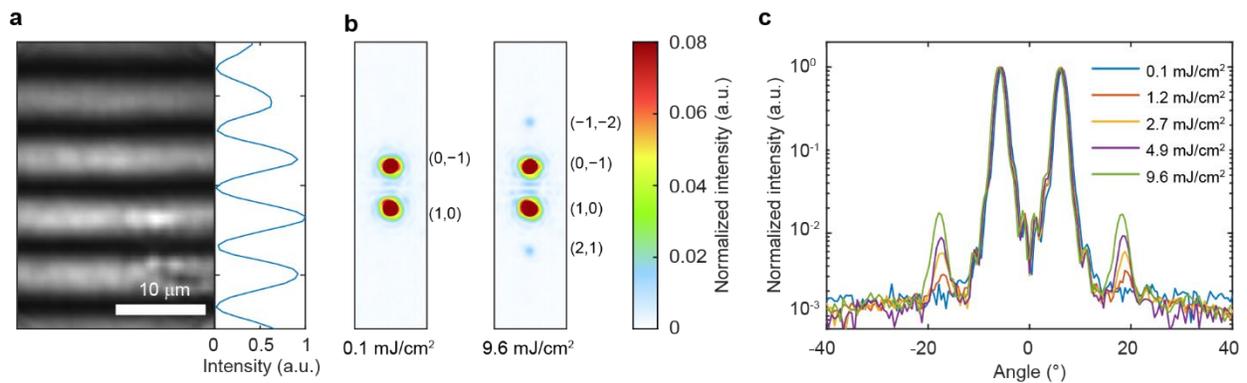

**Supplementary Figure 4 | Spatiotemporal self-diffraction on a transient optical grating with $\Lambda = 6.62$ μm. a**, Experimentally measured light intensity at the metasurface with an optical grating period of $\Lambda = 6.62$ μm. **b**, Experimentally measured Fourier plane images of the light transmitted through the metasurface at pump fluence of $F = 0.1$ mJ/cm$^2$ and of $F = 9.6$ mJ/cm$^2$ including the labeled grating orders. **c,** Line cross-section of the experimentally measured Fourier plane images of the light transmitted through the metasurface at varying pump fluence. For all panels, the metasurface with $L = 619$ nm, $H = 695$ nm, and $P = 736$ nm is resonantly illuminated.

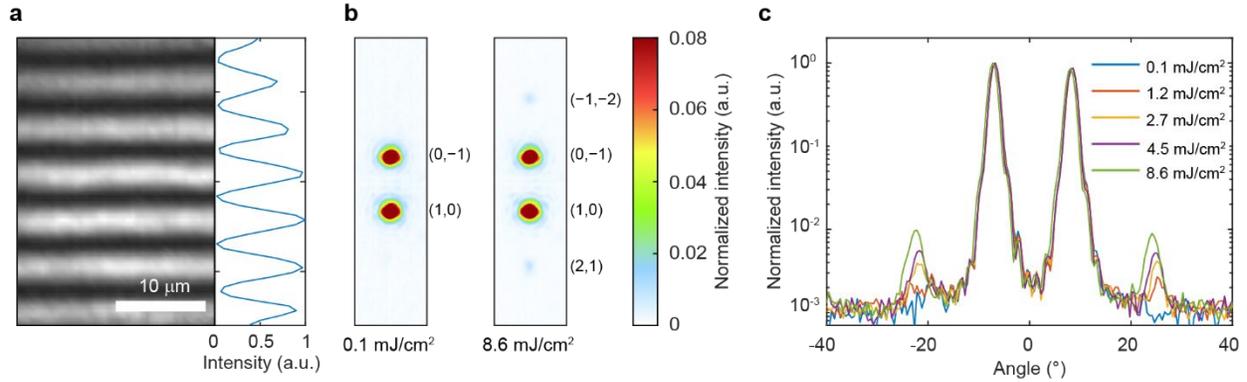

**Supplementary Figure 5 | Spatiotemporal self-diffraction on a transient optical grating with $\Lambda$ = 5.15 µm. a**, Experimentally measured light intensity at the metasurface with an optical grating period of $\Lambda$ = 5.15 µm. **b**, Experimentally measured Fourier plane images of the light transmitted through the metasurface at pump fluence of $F$ = 0.1 mJ/cm² and of $F$ = 8.6 mJ/cm² including the labeled grating orders. **c,** Line cross-section of the experimentally measured Fourier plane images of the light transmitted through the metasurface at varying pump fluence. For all panels, the metasurface with $L$ = 619 nm, $H$ = 695 nm, and $P$ = 736 nm is resonantly illuminated.

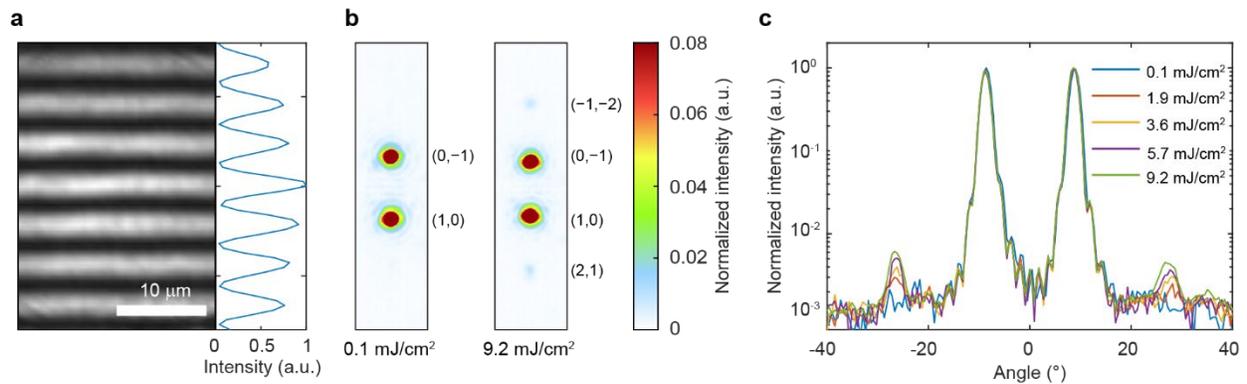

**Supplementary Figure 6 | Spatiotemporal self-diffraction on a transient optical grating with $\Lambda$ = 4.42 µm. a**, Experimentally measured light intensity at the metasurface with an optical grating period of $\Lambda$ = 4.42 µm. **b**, Experimentally measured Fourier plane images of the light transmitted through the metasurface at pump fluence of $F$ = 0.1 mJ/cm² and of $F$ = 9.2 mJ/cm² including the labeled grating orders. **c,** Line cross-section of the experimentally measured Fourier plane images of the light transmitted through the metasurface at varying pump fluence. For all panels, the metasurface with $L$ = 619 nm, $H$ = 695 nm, and $P$ = 736 nm is resonantly illuminated.

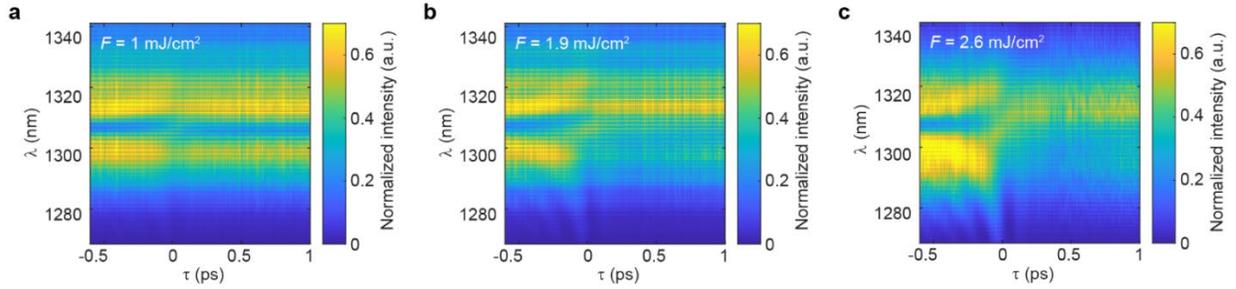

**Supplementary Figure 7 | Spectral analysis of transient pump-probe transmission measurements.** Experimentally measured normalized spectral transmitted intensity of the co-polarized probe beam with varying pump-probe delay times, for a uniform pump illumination at $\theta_i$ = 6.4° and a pump fluence of (**a**) $F$ = 1 mJ/cm², (**b**) $F$ = 1.9 mJ/cm² and (**c**) $F$ = 2.6 mJ/cm².

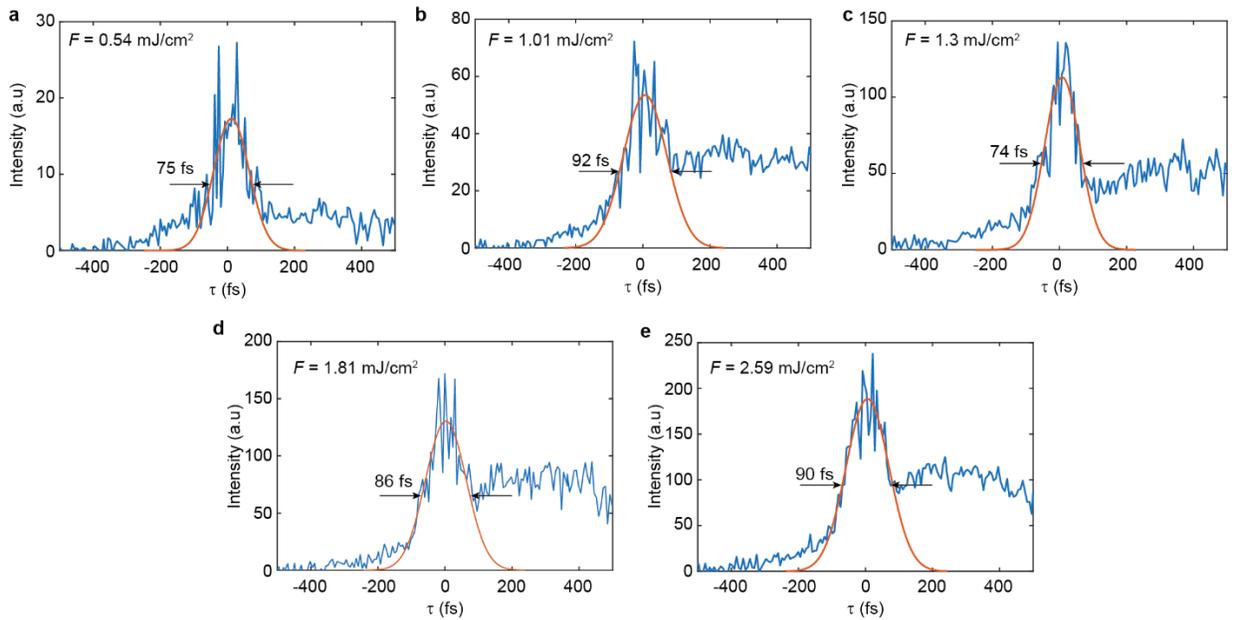

**Supplementary Figure 8 | Close-up time-dependent analysis of a metasurface for ultrafast all-optical beam deflection.** Experimentally measured normalized intensity of the probe beam transmitted to the (+2,+1) diffraction order with varying pump-probe delay times, for grating illumination at $\theta_i = \pm 6.4°$ and varying pump fluence of (**a**) $F$ = 0.54 mJ/cm², (**b**) $F$ = 1.01 mJ/cm², (**c**) $F$ = 1.3 mJ/cm², (**d**) $F$ = 1.81 mJ/cm², and (**e**) $F$ = 2.59 mJ/cm². This corresponds to a zoomed-in view of Fig. 4d. The numerical values in the figure denote the FWHM of a gaussian fit to the Kerr component of the deflected light intensity.

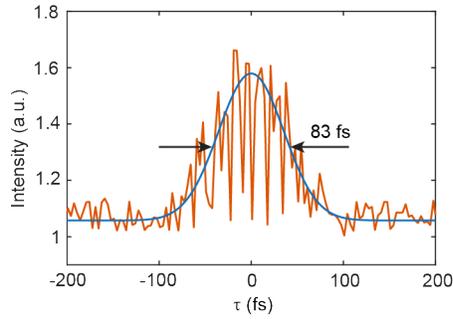

**Supplementary Figure 9 | Linear autocorrelation of the pump pulse.** Experimentally measured linear intensity autocorrelation of the pump pulse. Assuming two identical transform-limited pulses, a Gaussian fit to the autocorrelation suggests a pulse length of 83 fs. The effective pulse length of the pump is longer than that due to possible chirp and pulse dispersion in the optical measurement apparatus.

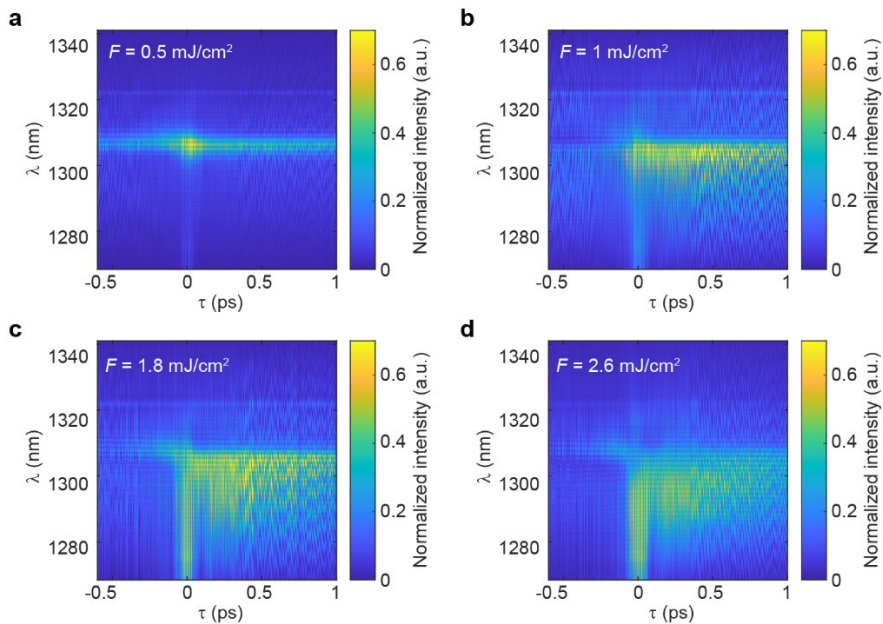

**Supplementary Figure 10 | Spectral analysis of ultrafast all-optical deflection of a co-polarized probe beam.** Experimentally measured normalized spectral intensity of the co-polarized probe beam transmitted to the (2,1) diffraction order with varying pump-probe delay times, for grating illumination at $\theta_i = \pm 6.4°$ and for a pump fluence of (**a**) $F = 0.5$ mJ/cm$^2$, (**b**) $F = 1$ mJ/cm$^2$, (**c**) $F = 1.8$ mJ/cm$^2$, (**d**) $F = 2.6$ mJ/cm$^2$.

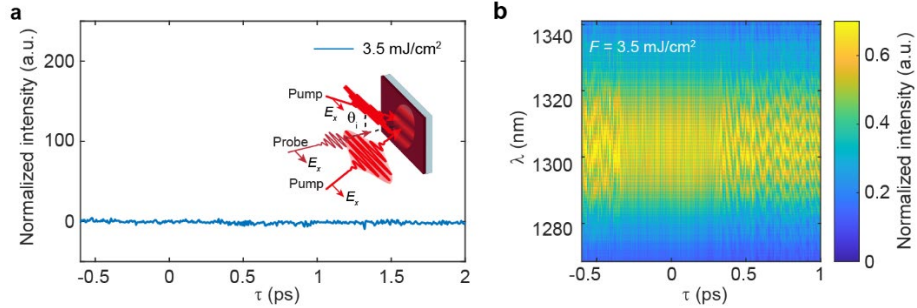

**Supplementary Figure 11 | Time-dependent analysis of a metasurface for ultrafast all-optical beam deflection on an a-Si/SiO$_2$ thin film stack. a,** Experimentally measured normalized intensity of the co-polarized probe beam transmitted to the (2,1) diffraction order with varying pump-probe time delays, for grating illumination at $\theta_i = \pm 6.4°$ and varying pump fluence. **b**, Spectral intensity of the light transmitted to the (2,1) diffraction order from the measurement in (**a**) at $F = 3.5$ mJ/cm$^2$ and varying pump-probe delay times. Pump and probe beams are co-polarized.

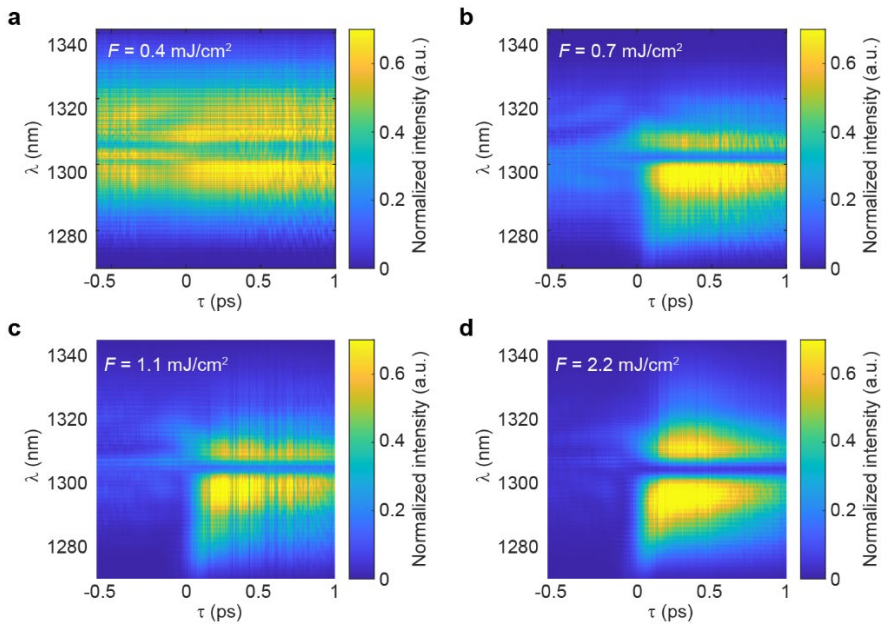

**Supplementary Figure 12 | Spectral analysis of ultrafast all-optical deflection of a cross-polarized probe beam.** Experimentally measured normalized spectral intensity of the cross-polarized probe beam transmitted to the (+2,+1) diffraction order with varying pump-probe delay times, for grating illumination at $\theta_i = \pm 6.4°$ and for a pump fluence of (**a**) $F = 0.4$ mJ/cm$^2$, (**b**) $F = 0.7$ mJ/cm$^2$, (**c**) $F = 1.1$ mJ/cm$^2$, (**d**) $F = 2.2$ mJ/cm$^2$.

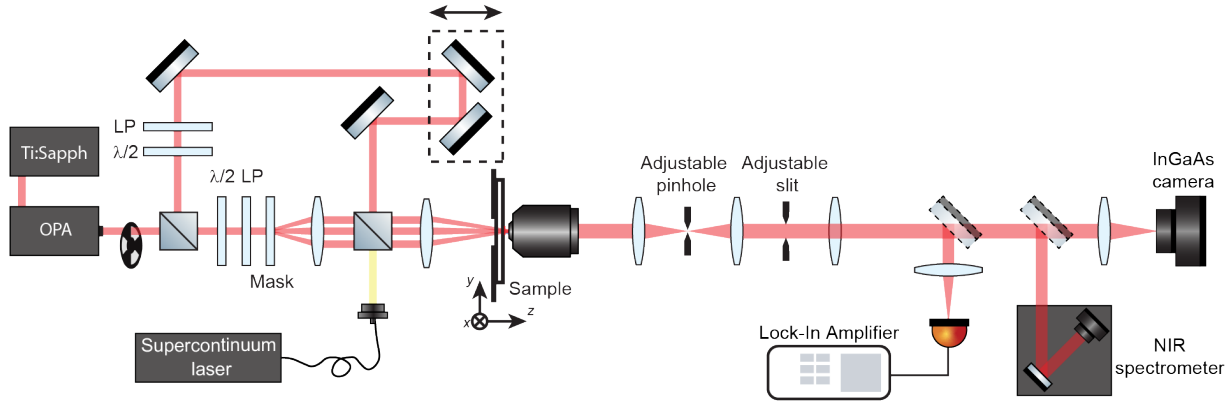

**Supplementary Figure 13 | Experimental setup.** For characterization, the metasurfaces are illuminated with loosely focused light from either a supercontinuum source or an optical parametric amplifier (OPA) fed by a Ti:Sapphire Amplifier. The laser pulses from the OPA are split into a pump and a probe beam, with the former modulated by a mask to create a range of pump illumination conditions. The transmitted light is collected by an objective lens (0.95 NA, 50x) and then sent to an InGaAs camera, photodiode, or grating spectrometer using the respective flip mirror. A set of half-wave plates ($\lambda/2$) and linear polarizers (LP) are used to set the polarization and intensity of the pump and probe beams. For Fourier plane imaging, a Fourier plane is formed on the camera sensor by exchanging the lens before the camera to a different focal length.

## Supplementary Note 1: Time-domain simulations

The time-dependent nonlinear optical properties of the metasurface were simulated using COMSOL Multiphysics. Specifically, a time-domain solver was used to solve the coupled Maxwell's equations, the rate equation of the charge carriers and the drift-diffusion equation of the charge carriers. For Maxwell's equations the following formulation was used:

$$\nabla \times \mu_r^{-1} (\nabla \times \mathbf{A}) + \mu_0 \sigma \frac{\partial \mathbf{A}}{\partial t} + \mu_0 \frac{\partial}{\partial t}\left(\epsilon_0 \epsilon_r \frac{\partial \mathbf{A}}{\partial t} - \mathbf{P}_{Kerr} - \mathbf{P}_{TPA}\right) + \mathbf{J}_{FC} = \mathbf{0}, \qquad (1)$$

where $\mathbf{A}$ denotes the magnetic vector potential, $\mathbf{J}_{FC}$ the free carrier density, and $\sigma$ the electric conductivity. $\mu_0$ and $\mu_r$ represent the vacuum and relative permeability, and $\epsilon_0$ and $\epsilon_r$ represent the vacuum and relative permittivity. The third-order nonlinear polarization due to the Kerr effect is described by

$$\mathbf{P}_{Kerr} = \epsilon_0 \chi^{(3)} \mathbf{E}^3, \qquad (2)$$

where $\chi^{(3)}$ = 2.45·10⁻¹⁹ m²V⁻² is the third order nonlinear susceptibility of amorphous silicon[1]. The nonlinear polarization due to two photon absorption is given by[2]

$$\mathbf{P}_{TPA} = -\frac{c^2 \epsilon_0^2 \epsilon_r \beta_{2PA}}{i\omega} \mathbf{E}^3, \qquad (3)$$

Where $c$ is the speed of light in vacuum and $\beta_{2PA}$ = 3.2 cm/GW is the two photon absorption coefficient of amorphous silicon[3]. The charge carrier density, $N_e$, in a-Si is modeled with the following equation[2,4]

$$\frac{\partial N_e}{\partial t} = \frac{N_a - N_e}{N_a} w_{2PA} + D_e \Delta N_e - \frac{N_e}{\tau_{rec}}, \qquad (4)$$

where $N_a$ corresponds to the atomic density, $D_e$ = 18 cm²/s to the carrier ambipolar diffusivity[5], and $\tau_{rec}$ = 6 ps to the carrier recombination time[6]. The two-photon ionization rate is given by[2]

$$w_{2PA} = \frac{c^2 \epsilon_0 n^2 \beta_{2PA}}{8\hbar\omega} |\mathbf{E}|^4. \qquad (5)$$

Finally, the free carrier current density is determined from the equation[4]

$$\frac{\partial \mathbf{J}_{FC}}{\partial t} = -\nu_e \mathbf{J}_{FC} + \frac{e^2 N_e(t)}{m_e^*} \mathbf{E}, \qquad (6)$$

where $m_e^*$ = 0.12 $m_e$ is the reduced electron mass and $\nu_e = 1/\tau_d = 1.25 \cdot 10^{15}$ Hz the electron collision frequency[7]. These coupled differential equations are solved using an implicit generalized alpha time-stepping method. A Drude correction to the permittivity due to the photogenerated carriers is applied as obtained by[8]

$$\Delta \epsilon_1 = \frac{-N_e e^2}{m_e^* \epsilon_0 (\omega^2 + \tau_d^{-2})}, \qquad (7)$$

and

$$\Delta \epsilon_2 = \frac{-\Delta \epsilon_1}{\omega \tau_d}, \qquad (8)$$

where $e$ corresponds to the elementary charge. While Eqs. (1)-(8) are solved for amorphous silicon, in glass and free space only Eq. (1) is retained without nonlinear polarization or current terms. The simulated geometry comprises an amorphous silicon trapezoidal prism of height $H$ = 700 nm, a top length of 563 nm, and bottom length of 643 nm, on a glass substrate with a 35 nm undercut, similar to what was previously analyzed[9]. The length of the top and bottom ends of the nanopillar were set to coincide with the resonance wavelength with the measurement. This is done to compensate for differences between simulated and fabricated geometry. All other dimensions were extracted from scanning electron microscope images.

Using this simulation formalism, we determined the transmission spectra of a Gaussian pulse resonant with the high-Q metasurface at different illumination powers (see Fig. 2e) and extracted the resonant wavelength of the transmission dip (see Fig. 2d). In addition to transmission spectra, the simulations also allow for observing the charge carrier density and current density in the metasurface unit cell at different time steps. For reference, in the simulations the excitation pulse is launched into the simulation domain at $t$ = 0 fs at a distance of 2 μm from the glass-air interface. Supplementary Figure 14 illustrates the simulated charge carrier density distribution and current density distribution in the metasurface unit cell recorded at t = 500 fs. Additionally, it is possible to locally monitor the temporal evolution of the refractive index as the light pulse impinges on and scatters from the metasurface. Supplementary Figure 15 shows the temporal evolution of the fluence-dependent total refractive index, and the components of the refractive index change that is due to the optical Kerr effect and the free carrier effect respectively. The refractive index is shown at the center of the a-Si nanopillar. A maximum total refractive index change of $\Delta n$ = 0.0153 is observed at 1 mJ/cm. A clear time lag between the refractive index change due to the Kerr effect and free carrier effect is observed. The refractive index change due to the Kerr effect peaks at $t$ = 288 fs and reaches a maximum of $\Delta n$ = 0.0176. The refractive index change due to the free carrier effect peaks at 367 fs and reaches a maximum of $\Delta n$ = -0.00436.

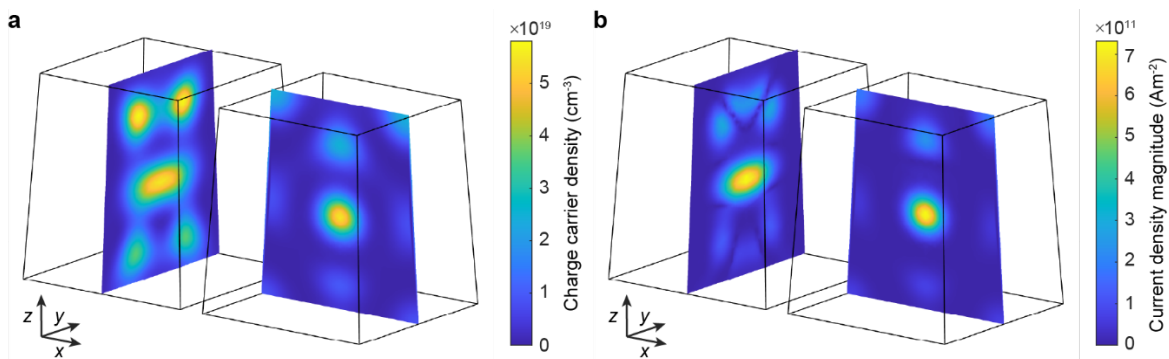

**Supplementary Figure 14 | Simulated charge carrier density and current density distribution in the metasurface. a**, Simulated charge carrier density distribution in the metasurface. **b,** Simulated current density distribution in the metasurface. The distributions are shown in the xz and yz planes at the midpoint of the nanopillar at $t$ = 500 fs, with the pulse being injected into the simulation domain at $t$ = 0 fs, from the substrate side of the metasurface. The pulse fluence in the simulation is $F$ = 1 mJ/cm$^2$.

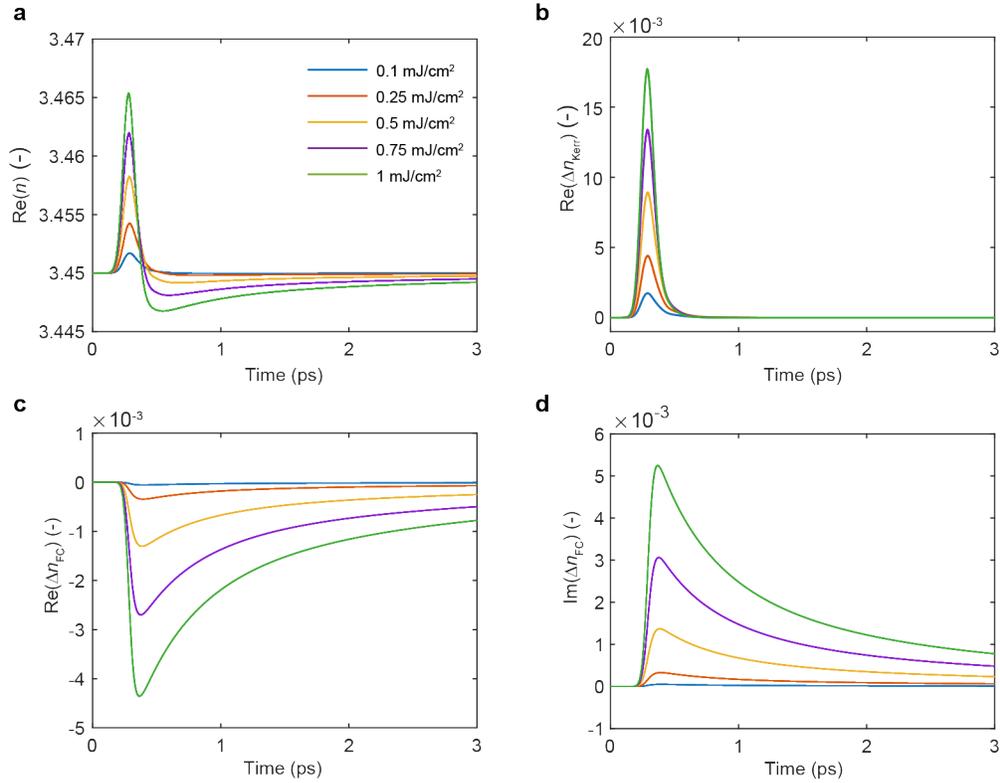

**Supplementary Figure 15 | Simulated transient refractive index changes of the metasurface. a**, Simulated transient power-dependent refractive index of the metasurface at the center point of a nanopillar. **b,** Simulated real part of refractive index change due to the optical Kerr effect. **c,** Simulated real part of the refractive index change due to photogenerated carriers. **d,** Simulated imaginary part of refractive index change due to photogenerated carriers. For reference, in the simulations the excitation pulse is launched into the simulation domain at t = 0 fs.